\documentclass[prd,aps,floatfix,nofootinbib,preprint]{revtex4}
\usepackage{latexsym}
\usepackage{graphicx}
\usepackage{amsmath}
\usepackage{hyperref}

\begin{document}
\hyphenation{fer-mio-nic per-tur-ba-tive pa-ra-me-tri-za-tion
pa-ra-me-tri-zed a-nom-al-ous}

\renewcommand{\thefootnote}{$*$}

\begin{center}
\vspace*{10mm}
{\large\bf Reply to: `Comment on: 
``\,'t Hooft vertices,
partial quenching, and rooted staggered QCD'''}
\\[12mm]
Claude~Bernard,$^a$ Maarten~Golterman,$^b$ Yigal~Shamir$^c$\ and\ 
Stephen~R.~Sharpe$^d$
\\[8mm]
{\small\it
$^a$Department of Physics
\\Washington University,
Saint Louis, MO 63130, USA}
\\[5mm]
{\small\it
$^b$Department of Physics and Astronomy
\\San Francisco State University,
San Francisco, CA 94132, USA}
\\[5mm]
{\small\it $^c$Raymond and Beverly Sackler School of Physics and Astronomy\\
Tel-Aviv University, Ramat~Aviv,~69978~Israel}
\\[5mm]
{\small\it $^d$Physics Department\\
University of Washington, Seattle, WA 98195, USA}
\\[10mm]
{ABSTRACT}
\\[2mm]
\end{center}

\begin{quotation}
We reply to Creutz's comments on our paper
``\,'t Hooft vertices, partial quenching, and rooted staggered QCD.''  
We show that his criticisms
are incorrect and result from a misunderstanding both of our work, and of
the related work of Adams. 
\end{quotation}

\renewcommand{\thefootnote}{\arabic{footnote}} \setcounter{footnote}{0}

\newpage

Creutz's fundamental criticism \cite{Creutz:2008kb}
of our work \cite{Bernard:2007eh}
is that we define (or act
as if we can define) the ``rooted continuum theory'' (RCT) in two 
inequivalent ways: by rooting four copies of a chirally invariant formulation, 
or by taking the continuum limit of rooted staggered quarks.  This is simply not true.
We define the RCT {\em only} in the former way, for which Creutz agrees
that the ``correctness of rooting is a trivial mathematical identity" \cite{Creutz:2008kb}.
Creutz therefore misses the whole point of our argument, which is that
the RCT serves as a counterexample to his ``proof'' \cite{CREUTZ}
  that rooted staggered
quarks are invalid.  The premises of the ``proof'' (in particular the
strong coupling between tastes induced by nonperturbative effects) apply
just as well to the RCT as to the actual rooted staggered theory.  
Yet in the RCT one can see by explicit, uncontroversial,
calculation that the conclusions of his argument do not hold:
there are no unphysical contributions to physical
correlation functions, and there is no problem constructing
single-taste observables.

As we have emphasized repeatedly \cite{Bernard:2007eh,Bernard:2006vv},
showing that Creutz's ``proof'' is incorrect is not the same as showing the validity
of rooted staggered  quarks. References \cite{Bernard:2007eh,Bernard:2006vv} do the
former, not the latter.  
The issue of the validity of rooted staggered quarks---{\it i.e.}, whether the known
lattice artifacts (nonunitarity and nonlocality
at physical scales when $a\not=0$ \cite{NON-UNITARY})  persist 
as $a\to 0$---is separate, and there is a
large body of analytic and numerical evidence \cite{REVIEWS} that
these artifacts vanish or decouple in the continuum limit
for any strictly positive quark mass(es).  Without  his
``proof,'' therefore, Creutz has no basis for his claim: ``The undesired effects 
[of rooted staggered fermions]\dots
will survive the continuum limit'' \cite{Creutz:2008kb}.

Recently, Adams \cite{Adams:2008db} has examined the rooting issue in
another simplified context, and also finds a counterexample to
Creutz's ``proof.'' It is therefore disingenuous for Creutz to
claim that ``rooting fails in this model also'' \cite{Creutz:2008kb} without mentioning
that Adams actually comes to the opposite conclusion.  We quote from the
abstract of Ref.~\cite{Adams:2008db}:
``Creutz's objections to the rooting trick apply just as much in this setting.
To counter them we show that the formulation has robust would-be zero-modes in topologically
nontrivial gauge backgrounds, and that these manifest themselves in a viable way in the rooted
fermion determinant and also in the disconnected piece of the pseudoscalar meson propagator as
required to solve the U(1) problem.''

\end{document}